# Microscopic analysis of p$^{10}$B scattering at the intermediate energies

A.V. Dzhazairov-Kahramanov[1*], E.T. Ibraeva[2†], and P.M. Krassovitskiy[2]
[1] *Al-Farabi Kazakh National University, 050040, av. Al-Farabi 71, Almaty, Kazakhstan*
[2] *Institute of nuclear physics RK, 050032, str. Ibragimova 1, Almaty, Kazakhstan*
[*] *albert-j@yandex.ru*
[†] *ibraeva.elena@gmail.com*

**Abstract:** The differential cross sections of the p$^{10}$B scattering at energies of 197, 600 and 1000 MeV have been calculated within the framework of the Glauber theory. The operator of multiple scattering takes into account the contributions of single and double collisions. The contributions from protons scattering on the nucleons of different shells have been estimated in the single-scattering cross-section. The comparison with the experiment at 197 MeV and with DWBA calculation showed the adequate description of cross-sections in the region of the front angles.

## 1. Introduction

The nuclear properties of boron (B) isotopes are studied well enough and their study is continued, since many of them are the main absorbers of neutrons and constructive element of the fusion reactors. In addition to fundamental nuclear research activities related to new sources of energy and the reactors of new-generation, their applications are also important: disposal of radioactive waste, radiation safety, and medical use of radiation sources.

Analysis of the structure and characteristics of $^{10}$B (along with other 1*p*-shell nuclei) is performed by various methods: the Monte Carlo (MC) [1,2], the antisymmetrized molecular dynamics (AMD) [3−5], the tensor-optimized shell model (TOSM) [6], the diffraction (Glauber) scattering [7−9], the coupled-channel (CC) [10], the distorted wave Born approximation (DWBA) [10,11]. The energy spectra, impulse and density distribution, the static characteristics (quadrupole and magnetic moments, neutron and proton radius) were measured and calculated in these papers.

The paper [1] provides MC quantum calculations of the ground and lower excited states of $^{10}$B (and other light nuclei up to $^{12}$C), using the two-body (Argonne (AV18)), three-body (Urban IX (U IX) and the Illinois (Il1−Il5) potentials. The calculations have been performed using the correlated many-body WF with the LS-coupling, with α-particle core and p-shell nucleons, with the relevant ($J^{\pi}$,$T$) quantum numbers for the lower states. After optimization these variational trial functions are used as the introductory Green's functions in the Monte Carlo (GFMC) calculations of energy. The Illinois three-nucleon potentials reproduce 10 states of $^{10}$B (and other light nuclei) with the mean-square deviation less than 900 keV. The authors of [1] note that "The most intriguing result was that we found the new models, correctly predicted $3^+$ main state of $^{10}$B for the first time".

In the papers of the last few years [3−6,11] it was found that the structure of the $^{10}$B ground state can be represented as the three-cluster configuration $^{10}$B = α−pn−α, where the proton and neutron pair forms the deuteron cluster. The states of $^{10}$B positive parity have been calculated within the framework of AMD [5] and TOSM [6], taken into account the effects of spin-orbit interaction on the energy spectra and pn correlations in the ground and excited states. The energy spectrum was calculated with two types of interactions $AV8'_{eff}$ and MN. It was shown that the best agreement with the experiment is achieved with two- (NN) and three- (NNN) partial chiral forces in the shell model without core (NCSM) [11]. It was found that the contribution of the three-partial forces is important for ordering of $^{10}B$ states, which significantly improves the agreement with the experimental data.

The short-range forces/correlations in He-Li-Be-B-C nuclei (including $^{10}B$) are studied in [12]. The energy spectra, impulse distributions and mean-square radii were calculated in the potential model with Argonne (AV18) and Illinois (Il7) interactions. The calculations were performed by the method of Green's functions Monk Carlo (GFMC). It was shown that the correlations induced by central and tensor components of the $v_{ij}$ potential, affect the nuclear properties such as distribution of neutron and proton pair density, pulse distribution of nucleons and nucleon pairs in nuclei, on the two-nucleon emission in the processes of high-energy scattering.

The measurements of the differential cross sections (DCS) and polarization characteristics: the vector analyzing power ($A_y$), the polarization ($P$), the polarization-transfer coefficients ($D_{NN}$, $D_{LL}$ and etc.) were performed in the Indiana University Cyclotron Facility. The polarized proton beam from the cyclotrons was accelerated to 197 MeV [10]. The measured characteristics are well reproduced in the calculations with an empirical effective nucleon-nucleon interaction in the Raynal DWBA98 program, including the coupled channels in the ground state and the $3^+ \leftrightarrow 4^+$ transition. At the angles $\theta$<35−40° the agreement with the experiment for DCS and Ay is observed, and it is better in the calculation by CC method than by DWBA. There is no agreement at larger angles. The data obtained in [10] have been analyzed in [11] in no-core shell model (NCSM), defined within the $(0+2)\hbar\omega$ single particle space and the MK3W shell model interaction. The structure of the ground and excited states of $^{10}B$ has been calculated. The resulting elastic scattering cross section is in good agreement with the experimental data, as well as the analyzing power, at least for small momentum transfers.

In our paper we use the Glauber diffraction theory [13], which microscopically describes the dynamics of interaction and sensitive to details of the nuclear structure at the intermediate and high energies. Thus, the measured (in the National Superconducting Cyclotron Laboratory, Michigan, USA) total cross sections at the energies from 15 to 53 MeV/nucleon on $^6Li$, $^7Be$, $^{10}B$, $^{9,10,11}C$, $^{12}N$, $^{13,15}O$, and $^{17}Ne$ nuclei on the Si target [7] were calculated in the Glauber theory. The calculation even in the optical limit of the Glauber theory with the Gaussian oscillatory densities is in good agreement with the data of total cross sections and rms radii $\langle r_p^2\rangle^{1/2}, \langle r_n^2\rangle^{1/2}, \langle r_m^2\rangle^{1/2}$, obtained from the scattering at high energies. The single-proton reactions of transfer were calculated in the extended Glauber model that includes the second order of noneikonal corrections, the realistic single-particle densities and spectrofactors of the shell model. Paper [9] provides the studied structural properties of some nuclei (Be-B-C-N and others), using the effective interaction in the framework of the microscopic non-relativistic Hartree-Fock and relativistic formalism of the mean field. To describe the dynamics of the reaction we also used the Glauber model with the densities obtained from these formalisms. A good agreement of the characteristics (binding energies, charge radii, quadrupole deformation parameters, densities, $\sigma_R$ reaction cross sections) with the experiment has been obtained.

The aim of our work is the microscopic description of DCS for elastic $p^{10}B$ scattering at the intermediate energies of 200-1000 MeV in the framework of Glauber theory considering single and double collisions in the operator $\Omega$ and comparing them with available data [10] and calculations in DWBA [11]. The partial contribution to the cross section of single scattering from collisions of protons with nucleons at different shells has been calculated.

## 2. Matrix element of $p^{10}B$ scattering

The amplitude of the elastic pA scattering in the Glauber approximation is defined as the integral of the matrix element $\sum_{M_J M_J'} \left\langle \Psi_f^{J'M_J'} \middle| \Omega \middle| \Psi_i^{JM_J} \right\rangle$ by the impact parameter $\vec{\rho}$, which is a two-dimensional vector in the plane perpendicular to the direction of the incident beam [13].

$$M_{if}(\vec{q}) = \sum_{M_J M_J'} \frac{ik}{2\pi} \int d^2\vec{\rho} \exp(i\vec{q}\vec{\rho}) \delta(\vec{R}_A) \left\langle \Psi_f^{J'M_J'} \left| \Omega \right| \Psi_i^{JM_J} \right\rangle , \tag{1}$$

where $\vec{R}_A$ is the coordinate in a center of mass (cm), $\Psi_i^{JM_J}, \Psi_f^{J'M_J'}$ are the wave functions (WFs) of the nucleus-target in the initial and final states, $\vec{k}, \vec{k}'$ – are the momenta of the projectile and ejected proton, $\vec{q}$ is the momentum transferred in the reaction: $\vec{q} = \vec{k} - \vec{k}'$. For elastic scattering $\left|\vec{k}\right| = \left|\vec{k}'\right|$, $q = 2k \sin\frac{\theta}{2}$.

The shell WF of $^{10}$B can be written as

$$\begin{aligned}\Psi_{if} &= \sum_{nlm} \Psi_{nlm} = \sum_i P_i \left| (1s)^4 (1p)^6 [f]^{(2T+1)(2S+1)} L_I \right\rangle = \\ &= P_{12} \left| (1s)^4 (1p)^6 [42]^{13} D \right\rangle + P_3 \left| (1s)^4 (1p)^6 [42]^{13} F \right\rangle = \\ &= \sum_m \left( P_{12} \varphi_{n=l=0}(\vec{r}_1,...\vec{r}_4) \varphi_{22m}(\vec{r}_5,...\vec{r}_{10}) + P_3 \varphi_{n=l=0}(\vec{r}_1,...\vec{r}_4) \varphi_{33m}(\vec{r}_5,...\vec{r}_{10}) \right),\end{aligned} \tag{2}$$

$$\varphi_{00}(\vec{r}_1,...\vec{r}_4) = \prod_{i=1}^{4} \varphi_{00}(\vec{r}_i); \quad \varphi_{22m}(\vec{r}_5,...\vec{r}_{10}) = \prod_{j=5}^{10} \varphi_{22m}(\vec{r}_j); \quad \varphi_{33m}(\vec{r}_5,...\vec{r}_{10}) = \prod_{j=5}^{10} \varphi_{33m}(\vec{r}_j); \tag{3}$$

$$\Psi_{if} = \Psi_{nlm}(\vec{r}) = P_{12} \prod_{i=1}^{4} \varphi_{00}(\vec{r}_i) \prod_{j=5}^{10} \varphi_{22m}(\vec{r}_j) + P_3 \prod_{i=1}^{4} \varphi_{00}(\vec{r}_i) \prod_{j=5}^{10} \varphi_{33m}(\vec{r}_j), \tag{4}$$

where $P_{12}, P_3$ are the weights of components [14].

The interaction of protons with nucleons of the nucleus is determined by the Glauber's operator of multiple scattering

$$\Omega = 1 - \prod_{i=1}^{A} \left(1 - \omega_i(\vec{\rho} - \vec{\rho}_i)\right) = \sum_{i=1}^{A} \omega_i - \sum_{i \langle j} \omega_i \omega_j + \sum_{i \langle j \langle k} \omega_i \omega_j \omega_k - ...(-1)^{A-1} \omega_1 \omega_2 ... \omega_A , \tag{5}$$

where $\vec{\rho}_\nu$ is the two-dimensional analogue of the three-dimensional single-particle coordinates of $\vec{r}_\nu$ nucleons. Here, the first term of the series is responsible for single collisions of particles; the second one is for double and etc. up to the last term responsible for A-multiple collisions. Expansion (5) provides a convenient way to establish the importance of terms of single collisions and collisions of higher orders. In our calculation in the Ω operator we will be limited by double scattering, because it is known [13] that the series (5) converges quickly and each subsequent term of the series contributes to the cross section by orders of lower previous one.

The profile function $\omega_i(\vec{\rho} - \vec{\rho}_i)$

$$\omega_i(\vec{\rho} - \vec{\rho}_i) = \frac{1}{(2\pi ik)} \int d\vec{q}_\nu \exp\left(-i\vec{q}_i(\vec{\rho} - \vec{\rho}_i)\right) f_{pN}(q_\nu) \tag{6}$$

is determined through the elementary $f_{pN}(q)$ amplitude, parameterized so that its coefficients has real physical meaning and it can be effectively applied in the multiple scattering formalism:

$$f_{pN}(q_\nu) = \frac{k\sigma_{pN}}{4\pi} \left(i + \varepsilon_{pN}\right) \exp\left(-\beta_{pN}^2 q_\nu^{\,2} / 2\right). \tag{7}$$

Here $\sigma_{pN}$ is the total cross section of proton scattering on the nucleon, $\varepsilon_{pN}$ is the ratio of the real part of the amplitude to the imaginary, $\beta_{pN}$ is the parameter of the amplitude cone slope. The parameters of elementary pN-amplitude are the input parameters of the theory and determined from the independent experiments. The summary of the amplitudes parameters at various energy values is provided in [15].

The substitution of the multiple scattering (5) series into the amplitude (1) and its further integration by the impact parameter $d\vec{\rho}$ and the pulses transmitted in each scattering event $d\vec{q}_i,\dots d\vec{q}_k$, lead to the following result (see conclusion in the Appendix):

$$\tilde{\Omega} = \tilde{\Omega}_1 + \tilde{\Omega}_2 = \frac{2\pi}{ik} f_{pN}(q) \sum_{i=1}^{10} \tilde{\omega}_i - \left( \frac{2\pi}{ik} f_{pN}\left(\frac{q}{2}\right)\right)^2 \sum_{i<j=1}^{10} \tilde{\omega}_i \tilde{\omega}_j + .., \tag{8}$$

where the operators with «~» sign indicate the result of integration by $d\vec{\rho}$ and $d\vec{q}_i,\dots d\vec{q}_k$:

$$\sum_{i=1}^{10} \tilde{\omega}_i = \sum_{i=1}^{10} \exp(i\vec{q}\vec{\rho}_i) , \tag{9}$$

$$\sum_{i=1}^{10} \tilde{\omega}_i \tilde{\omega}_j = \sum_{i<j=1}^{10} \exp\left( i\frac{\vec{q}}{2}(\rho_i + \rho_j)\right) \delta(\rho_i - \rho_j) . \tag{10}$$

Let us divide operators of single and double collisions by the terms applicable for the nucleons in different shells. The sum will be as follows for single collisions:

$$\sum_{i=1}^{10} \tilde{\omega}_i = \sum_{i=1}^{4} \tilde{\omega}_i + \sum_{j=5}^{10} \tilde{\omega}_j , \tag{11}$$

where the first term describes the collision with the nucleons in the *s*-shell, the second is for the nucleons in the *p*-shell. For double collisions the sum consists of collisions with nucleons in the *s*-, *p*- and *sp*-shells:

$$\sum_{i\prec j=1}^{10} \tilde{\omega}_i \tilde{\omega}_j = \sum_{i\prec j=1}^{4} \tilde{\omega}_i \tilde{\omega}_j + \sum_{i\prec j=5}^{10} \tilde{\omega}_i \tilde{\omega}_j + \sum_{i=1}^{4} \tilde{\omega}_i \sum_{j=5}^{10} \tilde{\omega}_j . \tag{12}$$

Let us write the matrix element of single scattering. We substitute the operator (9) and WF (2)−(4) in the formula (1)

$$M_{if}^{(1)}(\vec{q}) = \frac{k}{4\pi} f_{pN}(q) \left\langle \Psi_f \left| \sum \tilde{\omega}_i \right| \Psi_i \right\rangle =$$
$$= \frac{k}{4\pi} f_{pN}(q) \Big\{ P_{12}^2 N_{22} \left\langle \varphi_{00} \left| \sum_{i=1}^{4} \tilde{\omega}_i \right| \varphi_{00} \right\rangle + P_{12}^2 N_{00} \left\langle \varphi_{22m} \left| \sum_{j=5}^{10} \tilde{\omega}_j \right| \varphi_{22m} \right\rangle + P_3^2 N_{33} \left\langle \varphi_{00} \left| \sum_{i=1}^{4} \tilde{\omega}_i \right| \varphi_{00} \right\rangle +$$
$$+ P_3^2 N_{00} \left\langle \varphi_{33m'} \left| \sum_{j=5}^{10} \tilde{\omega}_i \right| \varphi_{33m} \right\rangle + P_3 P_{12} N_{00} \left( \left\langle \varphi_{33m'} \left| \sum_{j=5}^{10} \tilde{\omega}_j \right| \varphi_{22m} \right\rangle + \left\langle \varphi_{22m'} \left| \sum_{i=1}^{5} \tilde{\omega}_i \right| \varphi_{33m} \right\rangle \right) \Big\} \tag{13}$$

where $N_{00}$, $N_{22}$, $N_{33}$ are normalizations of WF

$$N_{00} = \left\langle \varphi_{00} \middle| \varphi_{00} \right\rangle = 1; \; N_{22} = \left\langle \varphi_{22m'} \middle| \varphi_{22m} \right\rangle = 1; \; N_{33} = \left\langle \varphi_{33m} \middle| \varphi_{33m} \right\rangle = 1 . \tag{14}$$

Let us identify the partial matrix elements

$$M_{00}^{(1)}(q) = \langle \varphi_{00} | \sum_{i=1}^{4} \widetilde{\omega}_i | \varphi_{00} \rangle, \tag{15}$$

$$M_{22}^{(1)}(q) = \langle \varphi_{22m'} | \sum_{j=5}^{10} \widetilde{\omega}_j | \varphi_{22m} \rangle, \tag{16}$$

$$M_{33}^{(1)}(q) = \langle \varphi_{33m'} | \sum_{j=5}^{10} \widetilde{\omega}_j | \varphi_{33m} \rangle, \tag{17}$$

$$M_{32}^{(1)}(q) + M_{23}^{(1)}(q) = \langle \varphi_{33m'} | \sum_{j=5}^{10} \widetilde{\omega}_j | \varphi_{22m} \rangle + \langle \varphi_{22m'} | \sum_{j=5}^{10} \widetilde{\omega}_j | \varphi_{33m} \rangle. \tag{18}$$

With new signs formula (13) can be written as:

$$M_{if}^{(1)}(\vec{q}) = \frac{k}{4\pi} f_{pN}(q) \{ P_{12}^2 M_{00}(\vec{q}) + P_{12}^2 M_{22}(\vec{q}) + P_3^2 M_{00}(\vec{q}) + P_3^2 M_{33}(\vec{q}) + P_3 P_{12} (M_{32}(\vec{q}) + M_{23}(\vec{q}) \}, \tag{19}$$

where $M_{00}(\vec{q})$ describes the collisions with nucleons of the *s*-shell, $M_{22}^{(1)}(\vec{q})$, $M_{33}$, $M_{32} + M_{23}$ describe the collisions with nucleons of the *p*-shell. The calculations of the matrix elements are given in the Appendix.

Write the matrix element of double scattering. The number of members in the operator $\sum_{i<j=1}^{10} \widetilde{\omega}_i \widetilde{\omega}_j$ (12) equals 45, including 6 members of scattering on the nucleons of the *s*-shell, 15 – on the nucleons of the *p*-shell, 24 – on the nucleons of the *sp*-shell. Let us substitute the operator (12) and WF (2)–(4) in the formula (1):

$$M_{if}^{(2)}(\vec{q}) = \frac{ik}{2\pi} \left( \frac{(2\pi)}{(ik)} f \left( \frac{q}{2} \right) \right)^2 \langle \Psi_f | \sum_{i \prec j} \widetilde{\omega}_i \widetilde{\omega}_j | \Psi_i \rangle =$$

$$= \frac{2\pi}{ik} f^2 \left( \frac{q}{2} \right) \{ (P_{12}^2 N_{22} + P_3^2 N_{33}) M_{00}^{(2)}(\vec{q}) + N_{00} \left( P_{12}^2 M_{22}^{(2)}(\vec{q}) + P_{33}^2 M_{33}^{(2)}(\vec{q}) + P_{12} P_3 \, (M_{23}^{(2)} + M_{32}^{(2)}) \right) +$$

$$+ M_{00}^{(1)}(\vec{q}) \left( (P_{22}^2 M_{22}^{(1)}(\vec{q}) + P_{22}^2 M_{33}^{(1)}(\vec{q}) + P_{12} P_3 \, (M_{23}^{(1)}(\vec{q}) + M_{32}^{(1)}(\vec{q}) \right) \}, \tag{20}$$

where

$$M_{00}^{(2)}(\vec{q}) = \langle \varphi_{00} | \sum_{i \prec j=1}^{4} \widetilde{\omega}_i \widetilde{\omega}_j | \varphi_{00} \rangle, \tag{21}$$

$$M_{22}^{(2)}(\vec{q}) = \langle \varphi_{22} | \sum_{i \prec j=5}^{10} \widetilde{\omega}_i \widetilde{\omega}_j | \varphi_{22} \rangle, \tag{22}$$

$$M_{33}^{(2)}(\vec{q}) = \langle \varphi_{33} | \sum_{i \prec j=5}^{10} \widetilde{\omega}_i \widetilde{\omega}_j | \varphi_{33} \rangle, \tag{23}$$

$$M_{23}^{(2)}(\vec{q}) = \langle \varphi_{22} | \sum_{i \prec j=5}^{10} \widetilde{\omega}_i \widetilde{\omega}_j | \varphi_{33} \rangle, \qquad M_{32}^{(2)}(\vec{q}) = \langle \varphi_{33} | \sum_{i \prec j=5}^{10} \widetilde{\omega}_i \widetilde{\omega}_j | \varphi_{22} \rangle \tag{24}$$

where $N_{00}, N_{22}, N_{33}$ are determined by formulae (14), and
$M_{00}^{(1)}(\vec{q}), M_{22}^{(1)}(\vec{q}), M_{33}^{(1)}(\vec{q}), M_{23}^{(1)}(\vec{q}) + M_{32}^{(1)}(\vec{q})$ – by formulae (16)–(18).

Furthermore, the calculations are made according to the scheme described for single scattering (see Appendix).

Having calculated the matrix elements, let us determine the DCS:

$$\frac{d\sigma}{d\Omega} = \frac{1}{2J+1}\left|M_{if}^{(1)}(\vec{q}) - M_{if}^{(2)}(\vec{q})\right|^2, \tag{25}$$

where the calculation of single collisions $M_{if}^{(1)}(\vec{q})$ was made according to formula (19), for double collisions $M_{if}^{(2)}(\vec{q})$ − according to formula (20). The “minus” sign appears because the series of multiple scattering (8) is sign-changing.

## 3. Discussion of results

We have calculated the DCS of $p^{10}B$ scattering in the Glauber diffraction theory in the approximation of the double scattering with WF in the shell model.

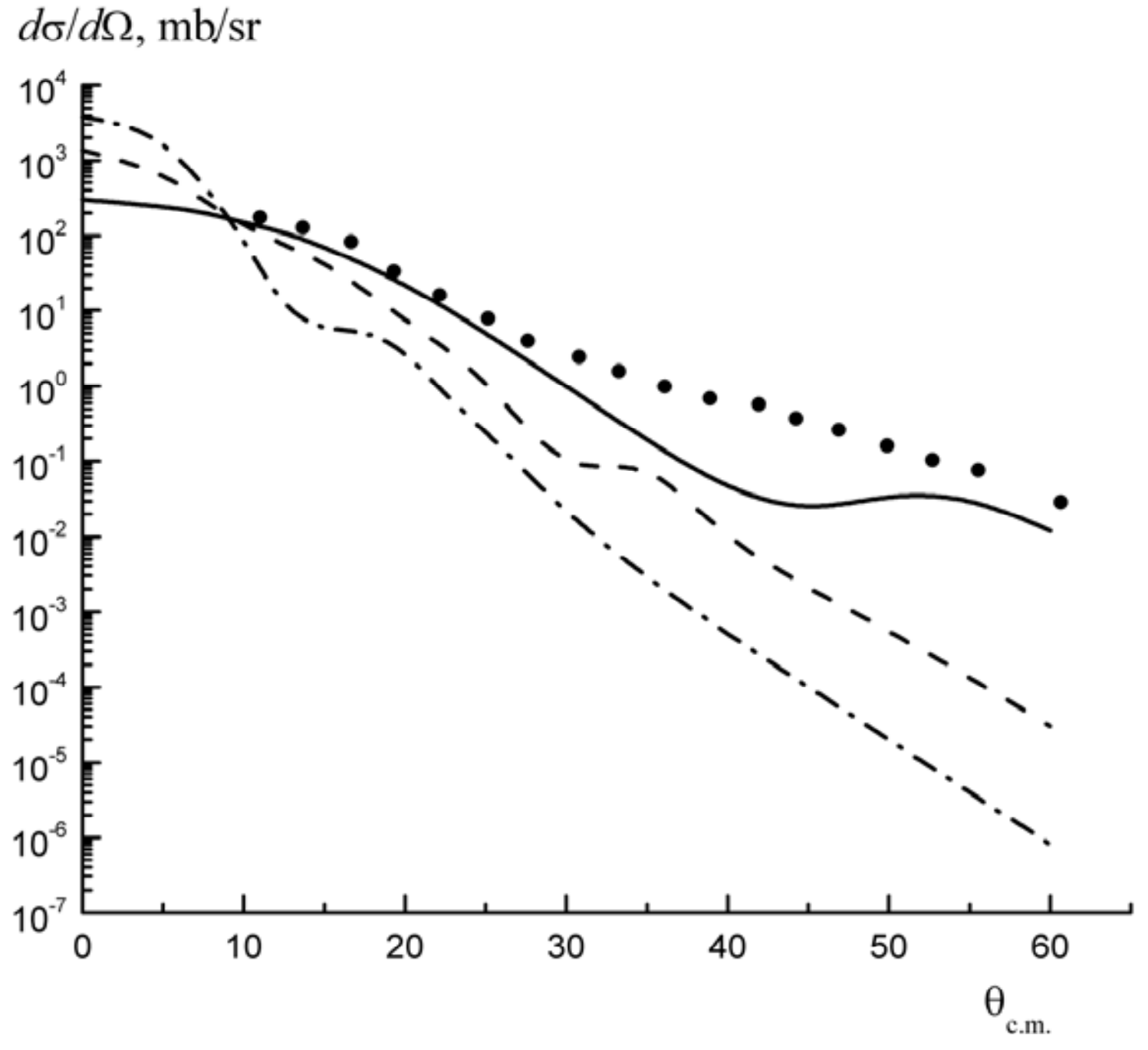

Fig. 1. The differential cross sections of protons scattering on $^{10}B$ at $E = 197$ (solid curve), 600 (dashed line) and 1000 (dashed-dotted line) MeV. Points are the experimental data from [10].

Fig. 1 shows the DCS of protons scattering on $^{10}B$ at different energies: 197 (solid line), 600 (dashed line) and 1000 (dashed-dotted line) MeV. The points are the experimental data obtained in the Indiana University Cyclotron Facility [10]. It can be seen that all cross sections decrease monotonically with increasing of the scattering angle. At zero angle the absolute value of DCS is greater at the higher energy (at E = 1000 MeV, it is an order of magnitude higher than at E = 197 MeV). In the case of energy increase, as the result of the scattering diffraction cone narrowing the diffraction minima in the DCS shift to the smaller angles region. So, the minimum is at $\theta \sim 42^o$ for E = 197 MeV, it is at $\theta \sim 30^o$ for E = 600 MeV and at $\theta \sim 15^o$ for E = 1000 MeV.

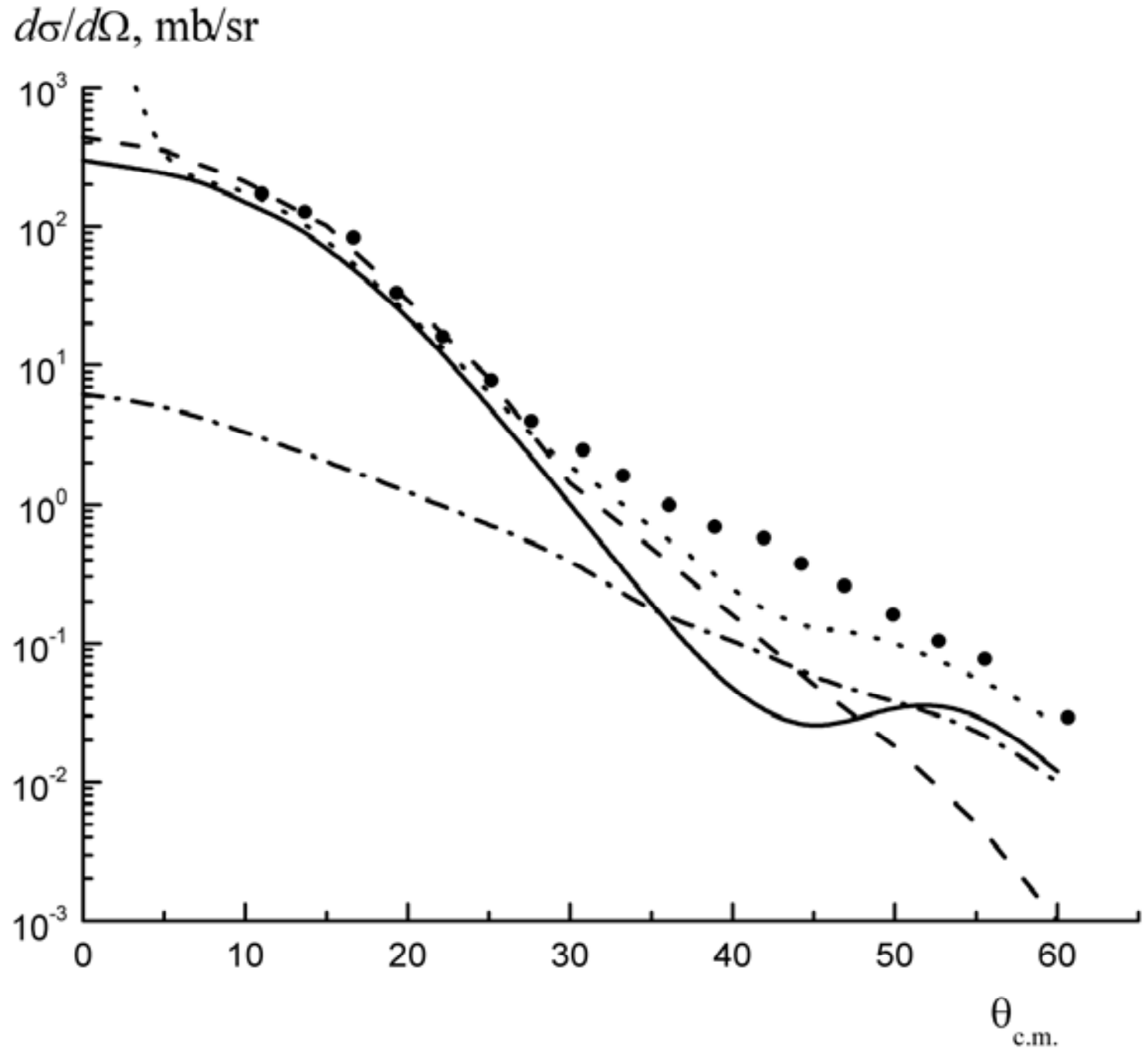

Fig. 2. Contribution of single and double collisions in DCS at E = 197 MeV. Dashed and dotted-dashed lines are the partial contributions of single and double collisions to the cross section, the solid lines are the total ones (the same as in Fig. 1). The dotted line is from [11].

Figs. 2-4 show the contribution of single and double collisions to the DCS. Here the dashed and the dashed-dotted lines are the partial contributions to the cross section of single and double collisions (formulae (19), (20)), the solid lines (formula (25) are the total ones (the same as in Fig. 1). The dotted line in Fig. 2 is taken from [11], where the calculation was carried out with the WF of the shell model without a core (LSSM) and in DWBA with the effective *NN*-interaction.

Comparison of the results of our calculation at 197 MeV with the experiment and the DWBA calculation shows that in the region of front angles $\theta < 30^o$ the cross section is well described in the approximation of the optical limit. At angles $55^o > \theta > 30^o$ the calculated curve is much lower than the experimental data. Here, we can observe the effect of extremely large quadrupole moment of $^{10}B$ ($Q$ = 84.72 mb). The filling effect of the diffraction minimum in the cross section from the contribution of the quadrupole scattering on the non-spherical nuclei was specified in the works [16,17]. The dotted line, calculated in the DWBA [11], describes the experiment best of all. However, it is a bit lower than the experimental data at $\theta \sim 43^o$.

From the figures it is clear that the contribution of single collisions dominates at small angles (at zero angle it is an order of magnitude greater than double one), but with the increase of the scattering angle it decreases rapidly and even at the angles ($\theta > 40°$ for 197 MeV, $\theta > 30°$ for 600 MeV and $\theta > 12°$ for 1000 MeV) the cross-section of double scattering is compared with it and further it gives the main contribution and determines the behavior of the DCS. The minimum appears in the crossing point of partial single and double cross sections. Thus, in the optical limit of the Glauber theory (when only single collisions are taken into account) we may describe the cross section only in the region of front angles, but already at angles $\theta > 35°$ we need to consider the double collisions. With the energy increase, the particles have more multiple collisions penetrating into the interior of the nucleus. As shown in [8,18,19] the contribution of triple scattering begins to affect the behavior of the cross sections at the angles $\theta > 40°–50°$, that is at the limit of the Glauber approximation.

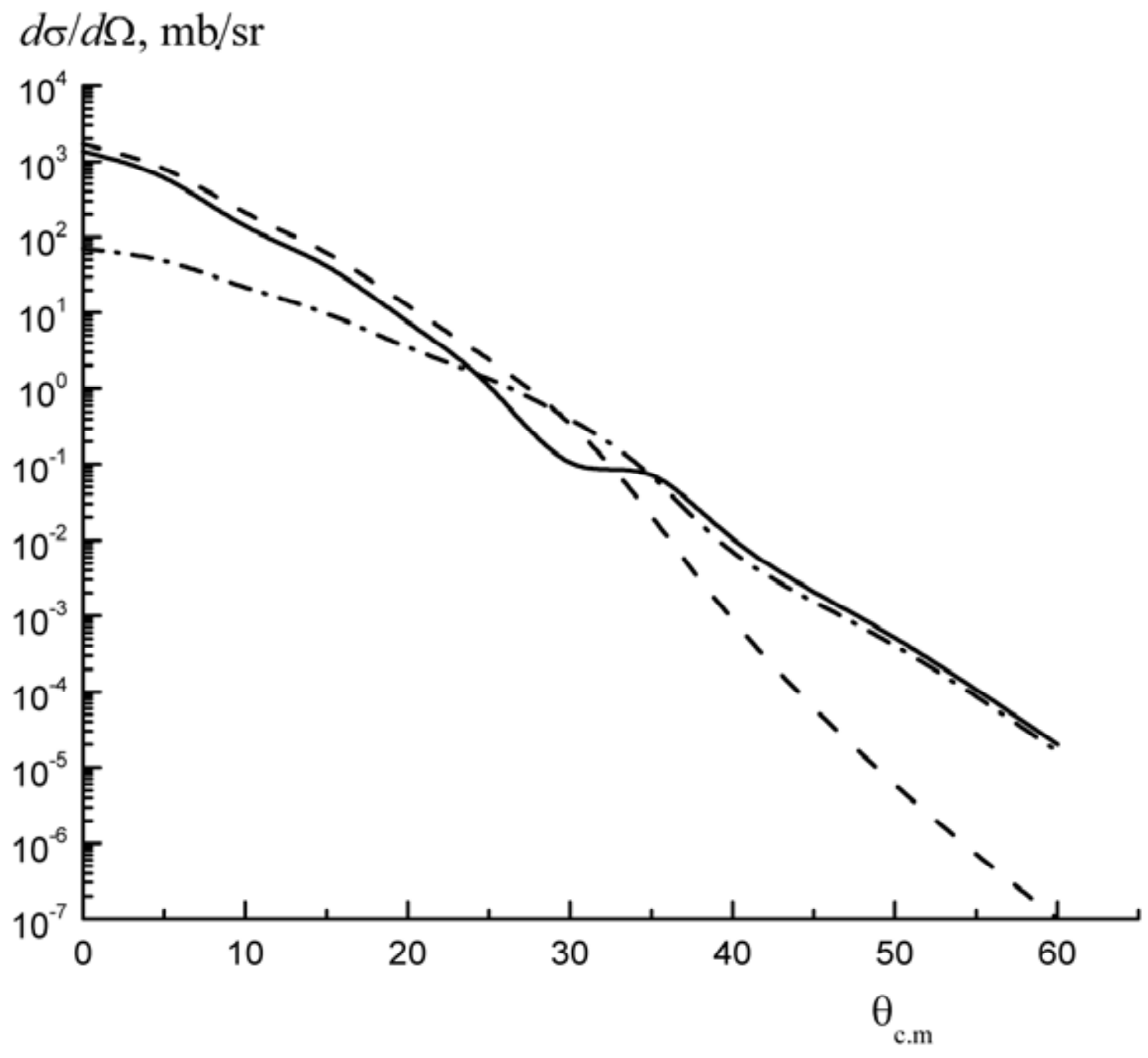

Fig. 3. Same as in Fig. 2 at E = 600 MeV.

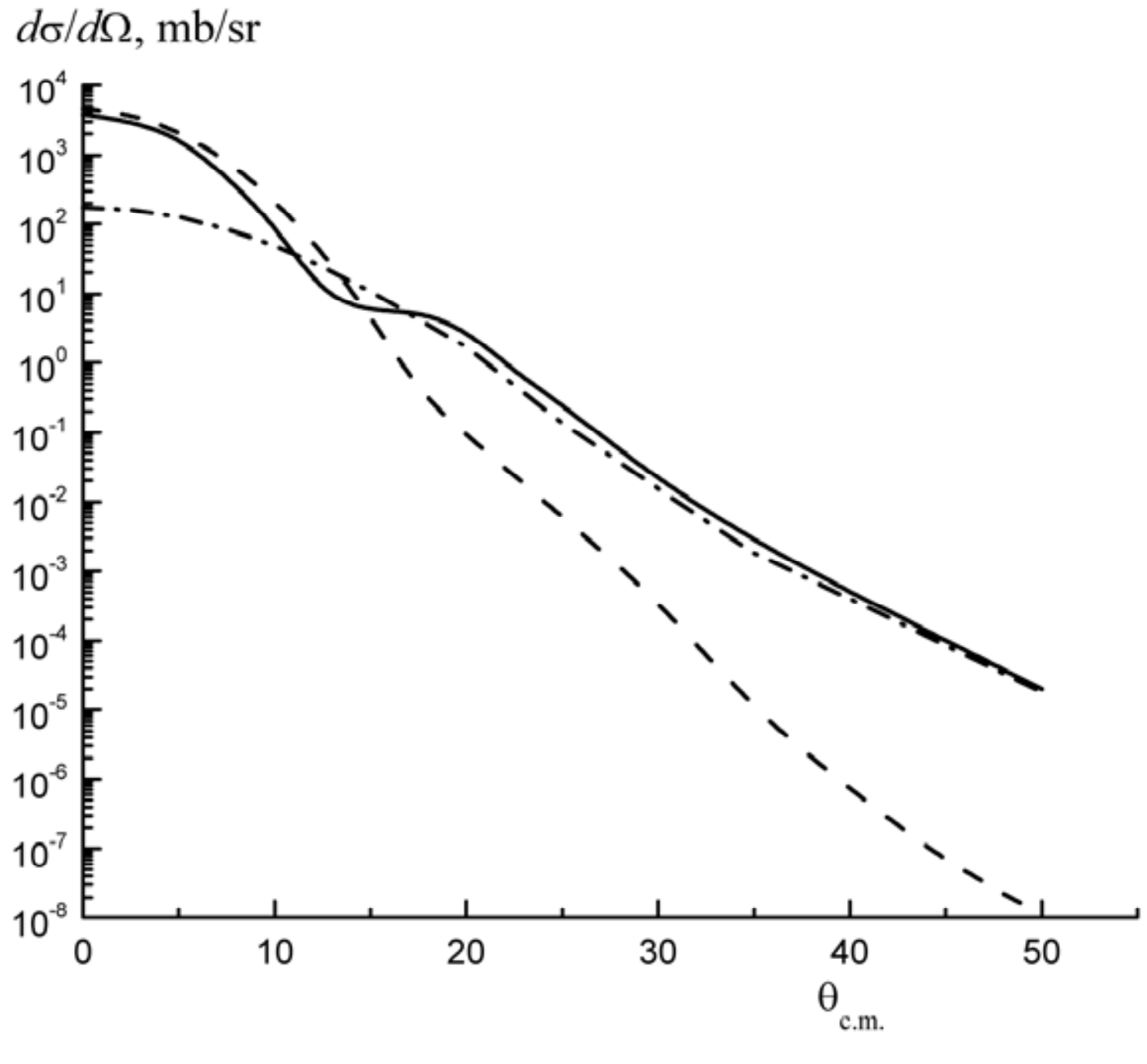

Fig. 4. Same as in Fig. 2 at E = 1000 MeV.

Figs. 5 and 6 show the partial cross sections of single collisions at proton scattering on nucleons from different shells at E = 197 MeV and 1000 MeV. The dashed, dashed-dotted and solid lines are contributions in the cross section from of scattering on nucleons of *s*-, *p*-shells and their total contribution (the same lines as the dashed ones in Figs. 2, 4.). At small angles (when the momentum transfer is small) the scattering occurs mainly on nucleons of the 1*p*-shell. At the momentum increase the protons begin to penetrate in the interior region of the nucleus and then the collisions with the nucleons of the *s*-shell become dominant. From the figures it is clear that at E = 1000 MeV the contribution from the partial cross sections of the scattering on the nucleons of the *s*-shell becomes determinative for smaller scattering angles.

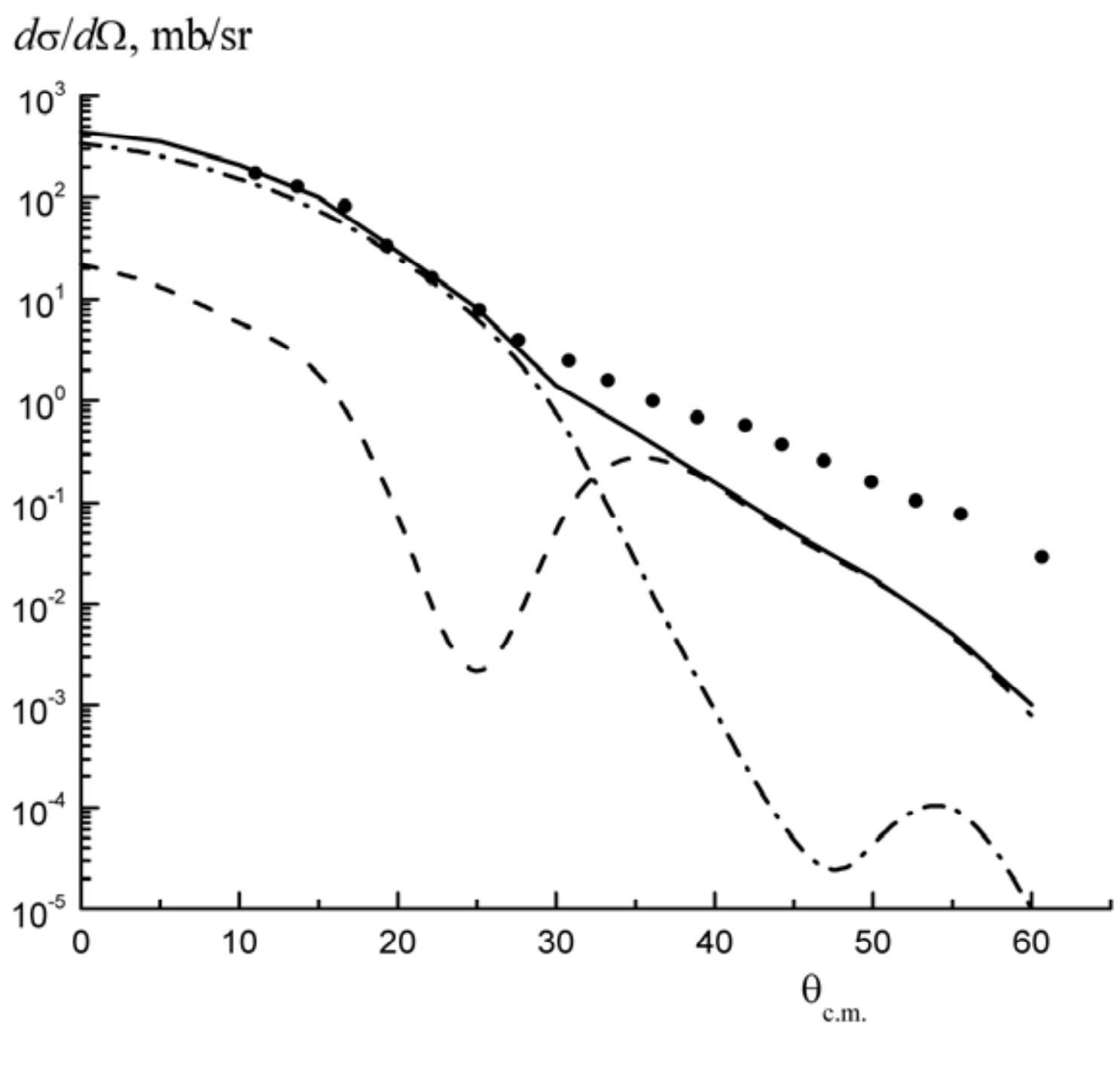

Fig. 5. Contributions of single collisions to the partial cross-section at proton scattering on nucleons from different shells at E = 0.197 GeV. Dashed, dashed-dotted and solid line are the contributions of scattering on nucleons of *s*-, *p*-shells to the cross section and their total contribution.

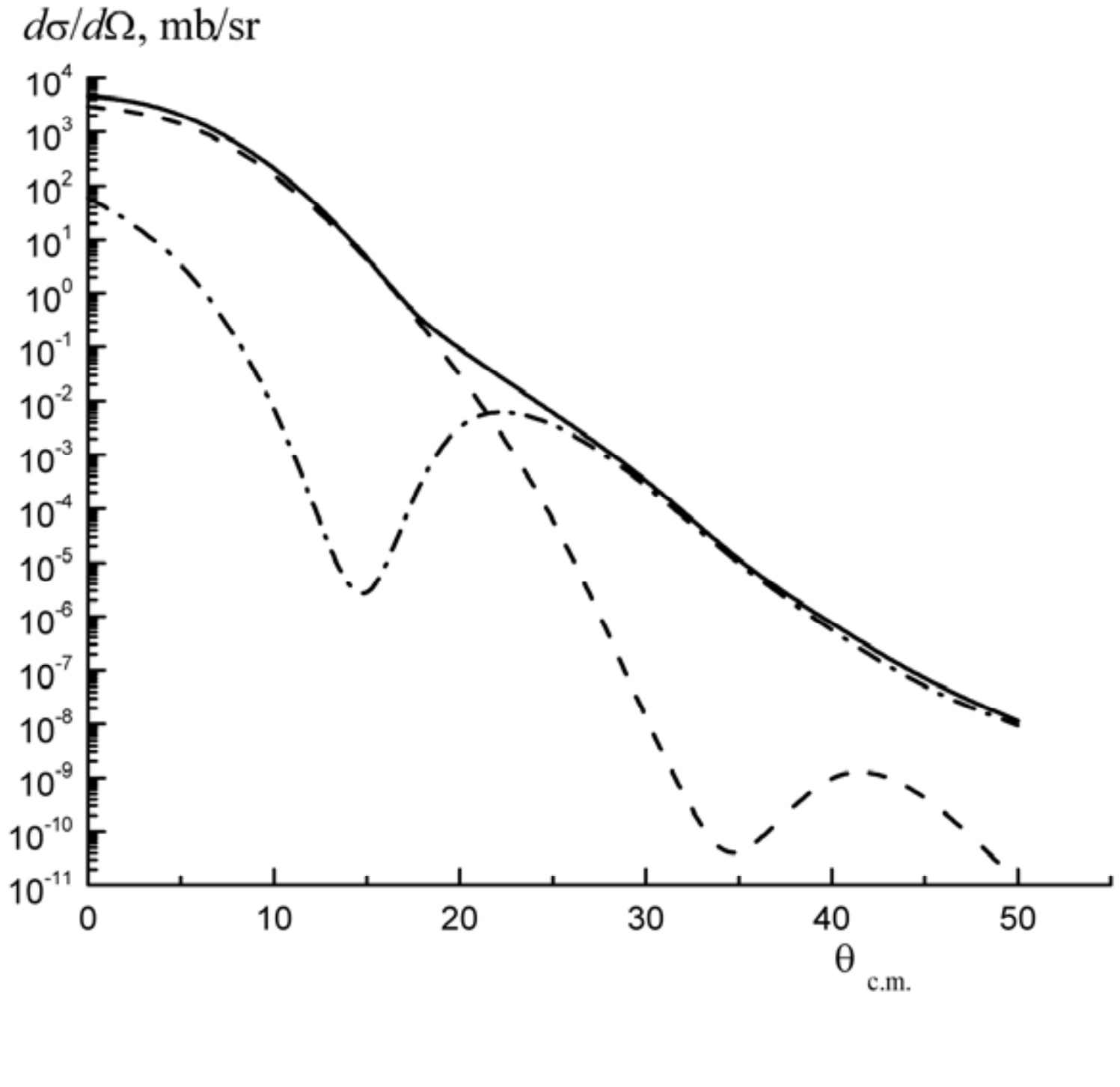

Fig. 6. Same as in Fig. 5 at E = 1000 MeV.

## 4. Conclusions

This paper calculates the DCSs for proton elastic scattering on $^{10}B$ at the energies E = 197, 600 and 1000 MeV. The derivation of matrix elements was made within the framework of the Glauber theory in the approximation of double scattering with the WF in the shell model. It enabled us to take into account the dynamics of the collisions and the structure of the WF and also to calculate the matrix elements analytically.

The comparison with the experiment at 197 MeV and with the DCS calculation in the DWBA showed that in the region of front angles (θ ~ 30º) our calculation agrees with the available data. The cross section of single scattering at zero angles is the order of magnitude larger than the cross section of double scattering, but in case of angle increase it decreases not so fast as single scattering and equals with it. It results in appearance of minimum specified by equality of single and double amplitudes that are parts of the DCS with opposite signs. The contribution of double collisions becomes dominant at higher momentum transfers/scattering angles. Accounting for the contribution of double collisions enables us to evaluate the optical limit of the Glauber approximation. Having compared the contribution of double collisions in the cross section at different energies we can see that their contribution increases with the energy increase. So, at E = 197 MeV the double collisions begin to exceed the single ones at θ > 40° and at E = 1000 MeV it occurs already at θ > 12°.

In the approximation of the optical limit (of single collisions) we took into account the scattering of protons on the nucleons of different shells. It is shown that the collisions with the nucleons of the *p*-shell defines the DCS at small momentum transfers (scattering angles) when the collisions occur in the surface region of the nucleus, but with the increase of momentum the nucleons penetrate into the interior region of the nucleus and the contribution from the scattering on the *s*-shell nucleons becomes dominant. Physically, this behavior of the DCS can be explained by the fact that the scattering on the nucleon in the internal 1*s*-shell requires greater impulse (than for the scattering on the nucleon of external 1*p*-shell), and the larger momentum needs the larger scattering angle.

## Acknowledgments

This work was performed under support of the program "Development of Nuclear Energy in the Republic of Kazakhstan".

## Appendix

Calculation of the matrix elements we will carry out in the spherical coordinate system. In the formula (9) let us replace the 2-dimensional vector of coordinate $\vec{\rho}_i$ by 3-dimensional one $\vec{r}_i$: $\vec{\rho}_i \to \vec{r}_i$ and expand $e^{i\vec{q}\vec{\rho}_i} \to e^{i\vec{q}\vec{r}_i}$ in the series of Bessel function and spherical functions

$$e^{i\vec{q}\vec{r}} = 4\pi \sum_{\lambda\mu} (i)^{\lambda} \sqrt{\frac{\pi}{2q}} J_{\lambda+\frac{l}{2}}(qr) Y_{\lambda\mu}(\hat{r}) Y_{\lambda\mu}(\hat{q}) \tag{A.1}$$

WF (4) is represented in the form of product of radial and spherical parts:

$$\varphi_{nlm}(\vec{r}) = R_{nl}(r) \cdot \mathrm{Y}_{lm}(\hat{r}), \tag{A.2}$$

where

$$R_{nl}\left(\frac{r}{r_0}\right)=\sqrt{\frac{2l+1}{\sqrt{\pi}r_0^3}}\exp\left(-\frac{r^2}{2r_0^2}\right)\sum_{k=0}^{\frac{n-l}{2}}\frac{(-1)^k}{k!}\left(\frac{r}{r_0}\right)^{2k+l}\sqrt{\frac{(n-l)!!(n+l+1)!!}{(n-l-2k)!!(2l+2k+1)!!}}, \tag{A.3}$$

$Y_{lm}(\hat{r})$ is the spherical function, $\hat{r}=\{\theta,\varphi\}$.

Let us calculate the partial matrix elements by substituting the operator (9) and WF (A.2) into formula (15).

$$M_{00}(q)=4\langle R_{00}(r)\mathrm{Y}_{00}(\hat{r})|\tilde{\omega}_1|R_{00}(r)\mathrm{Y}_{00}(\hat{r})\rangle=D\langle R_{00}\mathrm{Y}_{00}(\hat{r})|\sum_{\lambda\mu}\frac{1}{\sqrt{r}}J_{\lambda+\frac{1}{2}}(qr)\mathrm{Y}_{\lambda\mu}(\hat{r})\mathrm{Y}_{\lambda\mu}(\hat{q})|R_{00}\mathrm{Y}_{00}(\hat{r})\rangle=$$

$$=D\langle R_{00}(r)|\frac{1}{\sqrt{r}}J_{\frac{1}{2}}(qr)|R_{00}(r)\rangle\sum_{\lambda\mu}(i)^{\lambda}\langle\mathrm{Y}_{00}(\hat{r})|\mathrm{Y}_{\lambda\mu}(\hat{r})|\mathrm{Y}_{00}(\hat{r})\rangle\mathrm{Y}_{\lambda\mu}(q)=D\frac{1}{4\pi}B_{00}^{(1)}(q)\cdot \tag{A.4}$$

where

$$B_{00}^{(1)}(q)=\int_0^{\infty}|R_{00}|^2J_{\lambda+\frac{1}{2}}(qr)r^{3/2}dr, \tag{A.5}$$

$$D=16\pi\sqrt{\frac{\pi}{2q}} \tag{A.6}$$

$$M_{22}(\vec{q})=\langle\varphi_{22m'}|\sum_{j=5}^{10}\tilde{\omega}_j|\varphi_{22m}\rangle=\frac{3}{2}D\langle R_{22}\mathrm{Y}_{2m'}(\hat{r})|(i)^{\lambda}J_{\lambda+1/2}(qr)\mathrm{Y}_{\lambda\mu}(\hat{r})\mathrm{Y}_{\lambda\mu}(\hat{q})|R_{22}\mathrm{Y}_{2m}(\hat{r})\rangle=$$

$$=\frac{3}{2}D\langle R_{22}|J_{\lambda+\frac{1}{2}}(qr)\frac{1}{\sqrt{r}}|R_{22}\rangle\sum_{\lambda\mu}(i)^{\lambda}\langle\mathrm{Y}_{2m'}(\hat{r})|\mathrm{Y}_{\lambda\mu}(\hat{r})|\mathrm{Y}_{2m}(\hat{r})\rangle\cdot\mathrm{Y}_{\lambda\mu}(\hat{q})=\frac{3}{2}D\sum_{\lambda\mu}B_{22\lambda}^{(1)}\cdot F_{22\lambda}^{(2)}(\hat{q}) \tag{A.7}$$

$$B_{22\lambda}^{(1)}=\int_0^{\infty}\left|R_{22}{}^2\right|J_{\lambda+\frac{1}{2}}(qr)r^{3/2}dr \tag{A.8}$$

$$F_{22\lambda}^{(2)}(\hat{q})=\sum_{\lambda\mu b}(i)^{\lambda}\langle\mathrm{Y}_{2m'}(\hat{r})|\mathrm{Y}_{\lambda\mu}(\hat{r})|\mathrm{Y}_{2m}(\hat{r})\rangle\mathrm{Y}_{\lambda\mu}=\sum_{\lambda\mu m}(i)^{\lambda}\sqrt{2\lambda+1}\langle 20\lambda_0|20\rangle\langle 2m'\lambda\mu|2m\rangle\mathrm{Y}_{\lambda\mu}(\hat{q}) \tag{A.9}$$

After summation by $\lambda\mu m$ we have

$$F_{22\lambda=2}^{(2)}(\hat{q})=\mathrm{Y}_{00}+\frac{\sqrt{10}}{7}\left\{(2\sqrt{2}+\sqrt{3})(\mathrm{Y}_{22}(\hat{q})+\mathrm{Y}_{2-2}(\hat{q}))+\sqrt{3}(\mathrm{Y}_{21}(\hat{q})+\mathrm{Y}_{2-1}(\hat{q}))\right\} \tag{A.10}$$

$$M_{33}(\vec{q})=\langle\varphi_{33m'}|\sum_{j=1}^{5}\omega_j|\varphi_{33m}\rangle=6\cdot 4\pi\sqrt{\frac{\pi}{2q}}\langle R_{33}\mathrm{Y}_{3m'}(\hat{r})|\sum_{\lambda\mu}\frac{(i)^{\lambda}}{\sqrt{r}}J_{\lambda+\frac{1}{2}}(qr)\mathrm{Y}_{\lambda\mu}(\hat{r})\mathrm{Y}_{\lambda\mu}(\hat{q})|R_{33}\mathrm{Y}_{3m}(\hat{r})\rangle=$$

$$=\frac{3}{2}D\langle R_{33}|J_{\lambda+\frac{1}{2}}(qr)\frac{1}{\sqrt{r}}|R_{33}\rangle\sum_{\lambda\mu}(i)^{\lambda}\langle\mathrm{Y}_{3m'}(\hat{r})|\mathrm{Y}_{\lambda\mu}(\hat{r})|\mathrm{Y}_{3m}(\hat{r})\mathrm{Y}_{\lambda\mu}(\hat{q})\rangle=\frac{3}{2}D\sum_{\lambda\mu}B_{33\lambda}^{(1)}F_{33\lambda}^{(3)}(\hat{q}) \tag{A.11}$$

$$B_{33\lambda}^{(1)}=\int_0^{\infty}|R_{33}|^2J_{\lambda+\frac{1}{2}}(qr)r^{3/2}dr \tag{A.12}$$

$$F_{33\lambda}^{(3)}(\hat{q})=\sum_{\lambda\mu}(i)^{\lambda}\langle\mathrm{Y}_{3m'}(\hat{r})|\mathrm{Y}_{\lambda\mu}(\hat{r})|\mathrm{Y}_{3m}(\hat{r})\rangle\mathrm{Y}_{\lambda\mu}(\hat{q})=\sum_{\lambda\mu}(i)^{\lambda}\sqrt{2\lambda+1}\langle 30\lambda 0|30\rangle\langle 3m'\lambda\mu|3m\rangle\mathrm{Y}_{\lambda\mu}(\hat{q}) \tag{A.13}$$

After summation by $\lambda\mu m$ we have

$$F_{33}^{(3)}(\hat{q}) = \frac{7}{\sqrt{4\pi}} \mathrm{Y}_{00}\delta_{\lambda 0} - \sqrt{\frac{5}{3}}\left[\left(\frac{1}{\sqrt{2}}+1\right)\left(\mathrm{Y}_{22}(\hat{q}) + \mathrm{Y}_{2-2}(\hat{q})\right) + \left(\frac{\sqrt{5}}{2}+\frac{1}{10}\right)\left(\mathrm{Y}_{21}(\hat{q}) + \mathrm{Y}_{2-1}(\hat{q})\right)\right]\delta_{\lambda 2} \tag{A.14}$$

$$M_{23}(\vec{q}) = M_{32}(\vec{q}) = \langle \varphi_{22m'} | \sum_{j=5}^{10} \tilde{\omega}_j | \varphi_{33m} \rangle = \langle R_{22}\mathrm{Y}_{2m'}(\hat{r}) | \sum_{\lambda\mu} \frac{(i)^\lambda}{\sqrt{r}} J_{\lambda+\frac{1}{2}}(qr)\mathrm{Y}_{\lambda\mu}(\hat{r})\mathrm{Y}_{\lambda\mu}(\hat{q}) | R_{33}\mathrm{Y}_{3m}(\hat{r})\rangle =$$

$$= \langle R_{22}\mathrm{Y}_{2m'}(\hat{r}) | \sum_{\lambda\mu} \frac{(i)^\lambda}{\sqrt{r}} J_{\lambda+\frac{1}{2}}(qr)\mathrm{Y}_{\lambda\mu}(\hat{r})\mathrm{Y}_{\lambda\mu}(\hat{q}) | R_{33}\mathrm{Y}_{3m}(\hat{r})\rangle =$$

$$= \frac{3}{2} D \langle R_{22} | \frac{1}{\sqrt{r}} J_{\lambda+\frac{1}{2}}(qr) | R_{33}\rangle \sum_{\lambda\mu} (i)^\lambda \langle \mathrm{Y}_{2m'}(\hat{r}) | \mathrm{Y}_{\lambda\mu}(\hat{r}) | \mathrm{Y}_{3m}(\hat{r}) Y_{\lambda\mu}(\hat{q})\rangle = \frac{3}{2} D \sum_{\lambda} B_{23\lambda}^{(1)} F_{23\lambda}(\hat{q}) \tag{A.15}$$

where

$$B_{32\lambda} = \int_0^\infty R_{33} \cdot R_{22} J_{\lambda+\frac{1}{2}}(qr) \cdot r^{3/2} dr \tag{A.16}$$

$$\begin{aligned} F_{32\lambda}(\hat{q}) &= \sum_{\lambda} (i)^\lambda \langle \mathrm{Y}_{3m'}(\hat{r}) | \mathrm{Y}_{\lambda\mu}(\hat{r}) | \mathrm{Y}_{2m}(\hat{r})\rangle \mathrm{Y}_{\lambda\mu}(\hat{q}) = \\ &= \sum_{\lambda\mu} (i)^\lambda \sqrt{2\lambda+1} \langle 30\lambda 0 | 20\rangle \langle 3m'\lambda\mu | 2m\rangle \mathrm{Y}_{\lambda\mu}(\hat{q}) \end{aligned} \tag{A.17}$$

Consider the operator of double scattering (formula (6)). Let us integrate it by $d\vec{\rho}$ (substitution in formula (1)), then

$$\tilde{\Omega}_2 = \frac{1}{(2\pi i k)^2} \int d\vec{\rho}\, e^{i\vec{q}\vec{\rho}} \sum_{i \prec j=1}^{10} \omega_i \omega_j = \int d^2\rho\, e^{i(\vec{q}-\vec{q}_i-\vec{q}_j)\vec{\rho}} \int d\vec{q}_i d\vec{q}_j f_{pN}(q_i) f_{pN}(q_j)\, e^{i\vec{q}_i\vec{\rho}_i + \vec{q}_j\vec{\rho}_j} =$$

$$= (2\pi)^2 \int d\vec{q}_i\, d\vec{q}_j e^{i(\vec{q}_i\vec{\rho}_i + \vec{q}_j\vec{\rho}_j)} \delta(\vec{q} - \vec{q}_i - \vec{q}_j) f_{pN}(q_i) f_{pN}(q_j) \tag{A.18}$$

Let us introduce the new variables $\vec{\tilde{q}}_i = \vec{q}_i + \vec{q}_j$; $\vec{\tilde{q}}_j = \frac{1}{2}(\vec{q}_i - \vec{q}_j)$,

then $\vec{q}_i = \frac{\vec{\tilde{q}}_i}{2} + \vec{\tilde{q}}_j$, $\vec{q}_j = \frac{\vec{\tilde{q}}_i}{2} - \vec{\tilde{q}}_j$.

Having divided the variables in (A.18), neglecting the small impulse $\vec{\tilde{q}}_j$ in the elementary amplitude $f_{\rho N}$ and taking $f_{\rho N}$ outside the integral sign, after integrating by impulses $\vec{\tilde{q}}_{i,j}$ we have the following

$$\tilde{\Omega}_2 = \sum_{i<j} \tilde{\omega}_i \tilde{\omega}_j = \left(\frac{2\pi}{ik} f_{pN}\left(\frac{q}{2}\right)\right)^2 \sum_{i \prec j=1}^{10} e^{i\frac{\vec{q}}{2}(\vec{\rho}_i + \vec{\rho}_j)} \delta^2(\vec{\rho}_i - \vec{\rho}_j) = \left(\frac{2\pi}{ik} f_{pN}\left(\frac{q}{2}\right)\right)^2 \sum_{i=1}^{10} e^{i\vec{q}\vec{\rho}_i}\,. \tag{A.19}$$

The calculations of the matrix elements with the operator of double scattering are made according to the scheme given above.